\documentclass[10pt,showpacs]{revtex4}
\usepackage{graphicx,epsfig}

\def\xslash#1{{\rlap{$#1$}/}}
\def\half{\frac{1}{2}}
\def\beq{\begin{equation}}
\def\eeq{\end{equation}}
\def\beqa{\begin{eqnarray}}
\def\eeqa{\end{eqnarray}}
\def\iar{\begin{array}{l}}
\def\ear{\end{array}}

\begin{document}
\title{Renormalization of the Cabibbo-Kobayashi-Maskawa Matrix at One-Loop Level}
\author{Yong Zhou$^{a}$}
\affiliation{$^a$ Institute of Theoretical Physics, Chinese Academy of Sciences, 
         P.O. Box 2735, Beijing 100080, China}

\begin{abstract}
We have investigated the present renormalization prescriptions of the
Cabibbo-Kobayashi-Maskawa (CKM) quark mixing matrix. Based on one prescription 
which is formulated by comparing with the fictitious case of no mixing of quark generations, we propose a new prescription which can make the physical amplitude involving quark's mixing gauge independent and ultraviolet finite.
Compared with the previous prescriptions this prescription is very simple 
and suitable for actual calculations. Through analytical calculations we 
also give a strong Proof for the important hypothesis that in order to keep the 
CKM matrix gauge independent the unitarity of the CKM matrix must be preserved.
\end{abstract}

\maketitle

\section{Introduction}

Because of the difference between quark's mass eigenstates and its electroweak 
eigenstates, the Cabibbo-Kobayashi-Maskawa (CKM) quark's mixing matrix must be 
introduced in the standard model \cite{c1}. Although the effects of the renormalization
of CKM matrix is small due to the unitarity of the CKM matrix (GIM mechanism
\cite{c2}) - if effects of the quark masses are neglected, it is an theoretically important question in standard model. Along with the accurate measurement
of the CKM matrix elements develops very quickly \cite{c3}, the renormalization 
of CKM matrix becomes more important at present. 
Until now many people have discussed this problem using different methods 
\cite{c4,c5,c6,c7,c8,c9}. The early prescription constructed the CKM counterterm
by quark's wave-function renormalization constants (WRC) \cite{c6}. But 
recent calculation has shown that this prescription leads to the physical 
amplitudes involving quark's mixing gauge-dependent \cite{c7,c8}. Another 
kind of prescription is to renormalize the CKM matrix by comparing with 
the fictitious case of no quark's mixing \cite{c9,c4}. But Ref.\cite{c4} 
has pointed out that such prescription will break up the unitarity of the 
CKM matrix. Although a revised prescription to keep the unitarity of the CKM matrix 
\cite{c4} is present, it is very complex and unsuitable for actual 
calculations. So we propose a new prescription which is very simple and 
suitable for actual calculations compared with the previous prescriptions.
This is discussed in section 2. In section 3 we have discussed the relationship 
between the unitarity and the gauge independence of the CKM matrix through 
analytical calculations. Lastly we give our conclusions.

\section{One-Loop CKM Matrix Renormalization}

The gist of Ref.\cite{c9,c4} is to renormalize the transition amplitude 
of $W$ gauge boson decaying into two quarks in proportion to the same 
amplitude except for eliminating the quark-mixing effects, since such an 
amplitude is gauge independent and ultraviolet finite without introducing 
CKM matrix renormalization. At one-loop level the decay amplitude of 
$W^{+}\rightarrow u_i \bar{d}_j$ is \cite{c9}
\beqa
  T_1\,=\,&&A_L[V_{ij}(F_L+\frac{\delta g}{g}+\half\delta Z_W +
  \half\delta \bar{Z}^{uL}_{ii}+\half\delta Z^{dL}_{jj})+
  \sum_{k\not=i}\half\delta \bar{Z}^{uL}_{ik}V_{kj}+
  \sum_{k\not=j}\half V_{ik}\delta Z^{dL}_{kj}+\delta V_{ij}]
  \nonumber \\ &&+ 
  V_{ij}[A_R F_R+B_L G_L+B_R G_R]
\eeqa
with $g$ and $\delta g$ the $SU(2)$ coupling constant and its counterterm, 
$\delta Z_W$ the $W$ boson WRC, $\delta\bar{Z}^{uL}$ and $\delta Z^{dL}$ 
the left-handed up-type and down-type quark's WRC \cite{c10}, and
\beqa 
  A_L\,=\,&&\frac{g}{\sqrt{2}}\bar{u}_i(p_1){\xslash \varepsilon}\gamma_L 
  \nu_j(q-p_1) \nonumber \\
  B_L\,=\,&&\frac{g}{\sqrt{2}}\bar{u}_i(p_1)\frac{\varepsilon\cdot p_1}{M_W}
  \gamma_L \nu_j(q-p_1)
\eeqa
with $\varepsilon^{\mu}$ the $W$ boson polarization vector, $\gamma_L$ and 
$\gamma_R$ the left-handed and right-handed chiral operators, and $M_W$ the
$W$ boson mass. Similarly, replacing $\gamma_L$ by $\gamma_R$ in Eqs.(2) we
define $A_R$ and $B_R$, respectively. $F_{L,R}$ and $G_{L,R}$ are four form 
factors which come from the contributions of the irreducible electroweak 
one-loop diagrams for $W^{+}\rightarrow u_i \bar{d}_j$ (see Fig.1). 
\begin{figure}[htbp]
  \begin{center}
    \epsfig{file=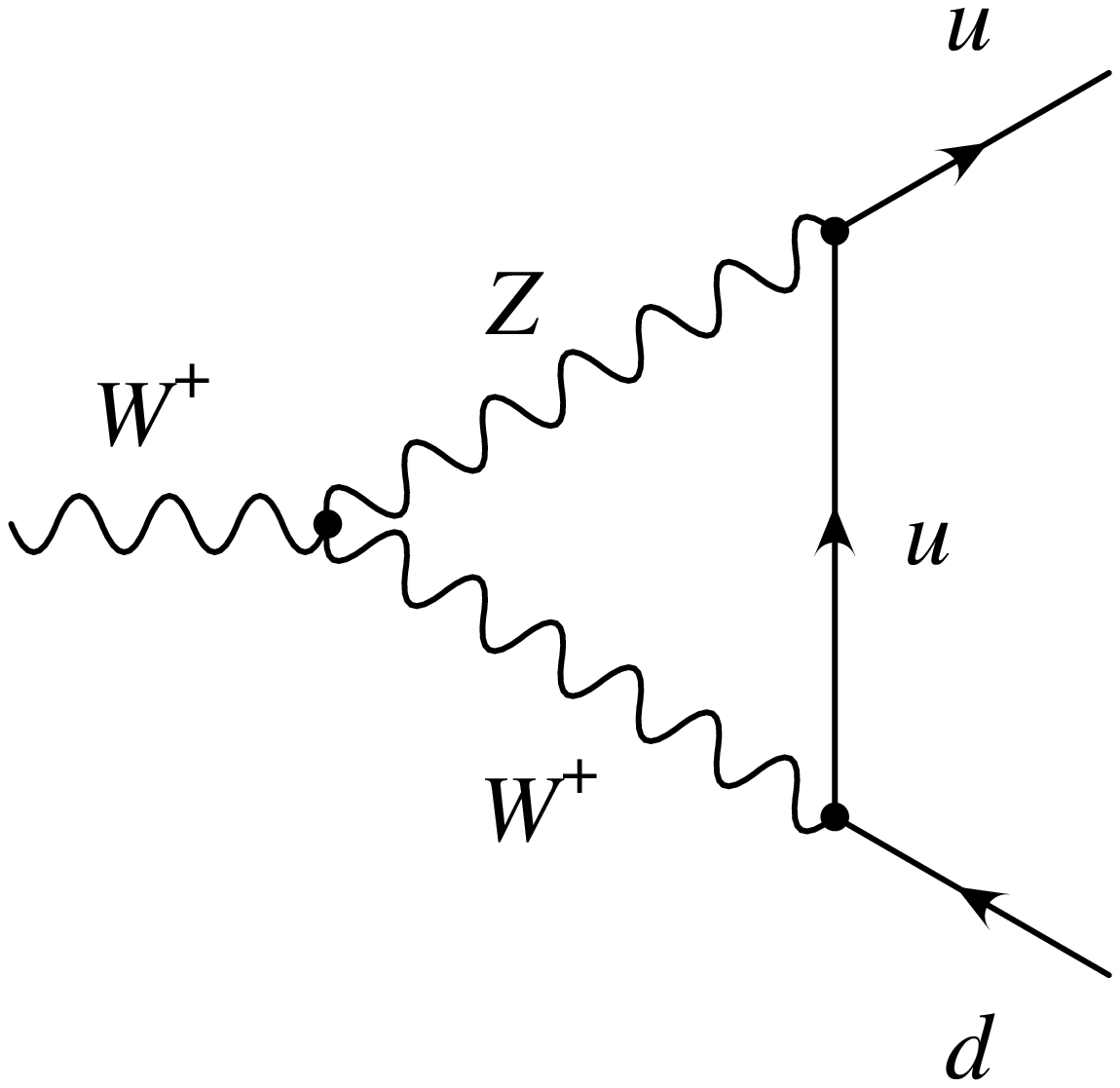,width=3.6cm}
    \epsfig{file=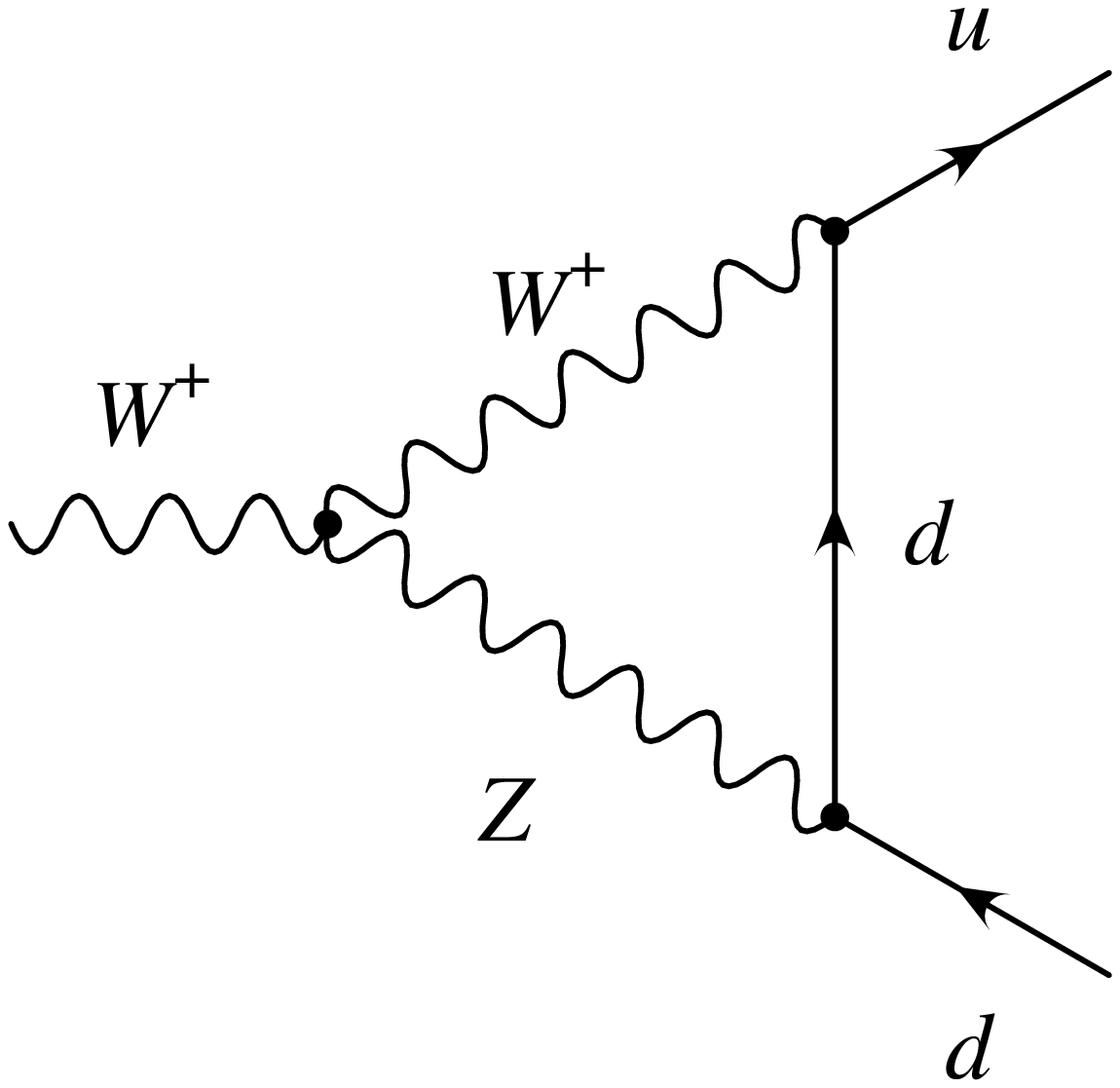,width=3.6cm}
    \epsfig{file=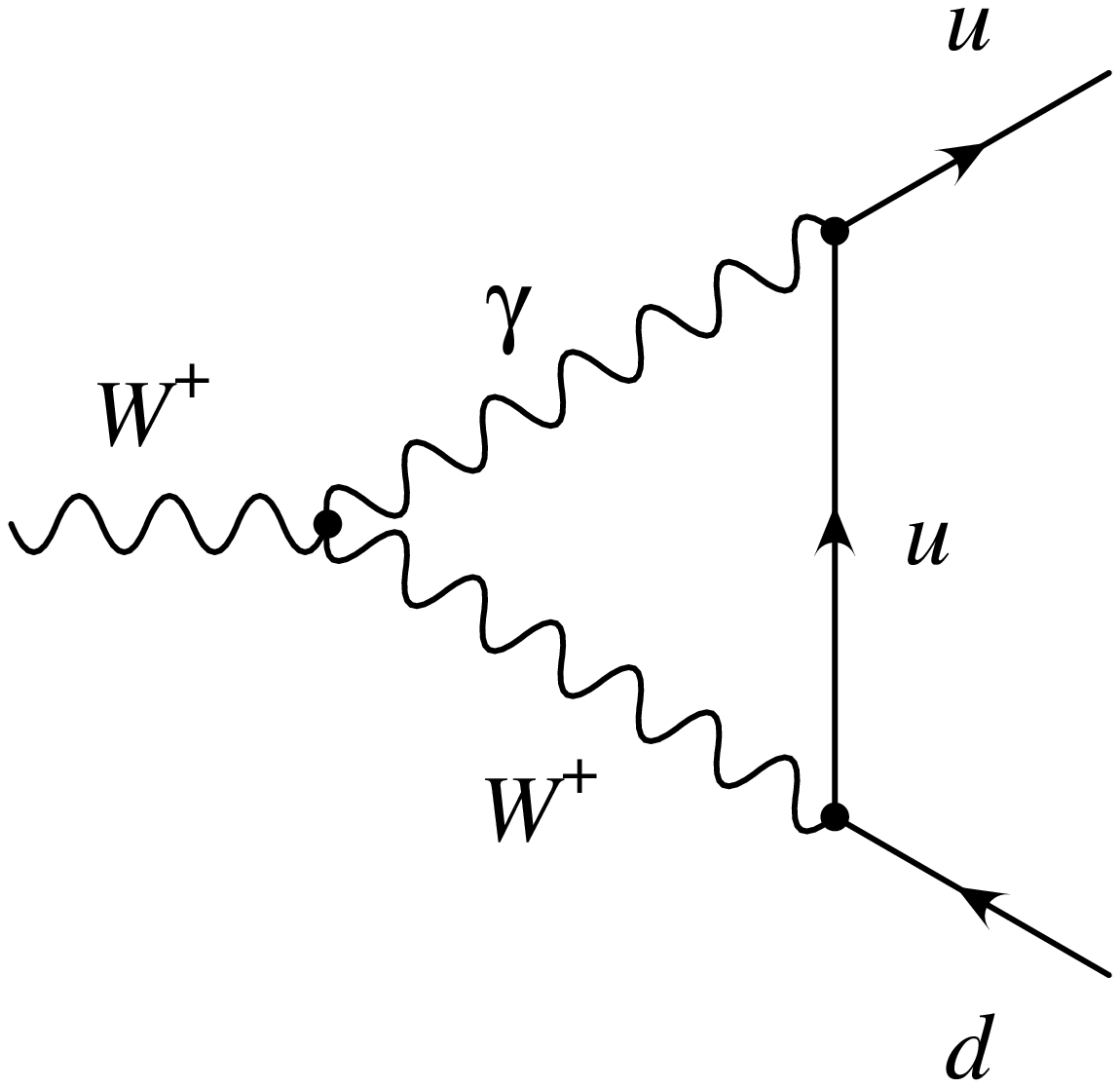,width=3.6cm}
    \epsfig{file=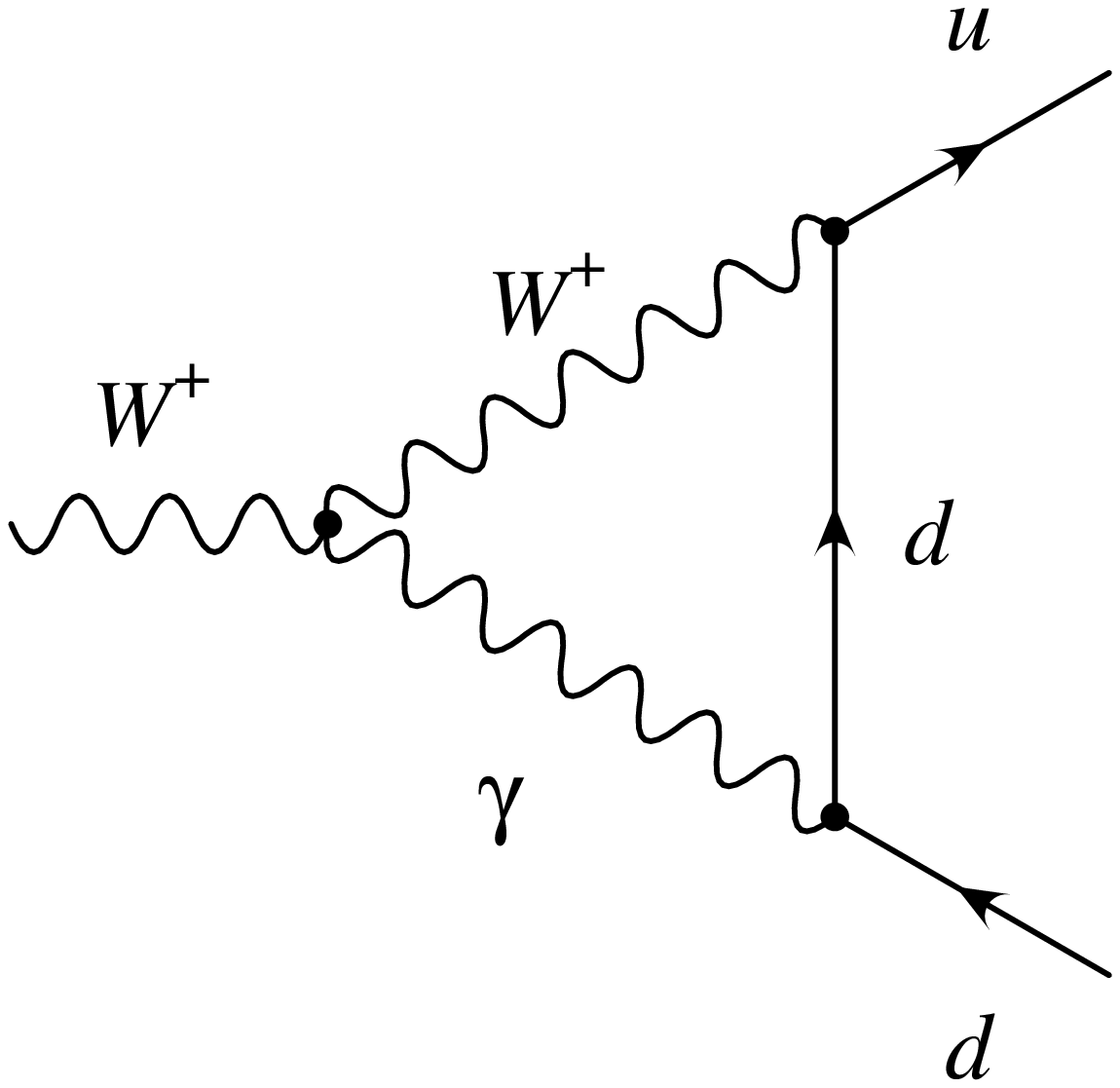,width=3.6cm}\\
    \epsfig{file=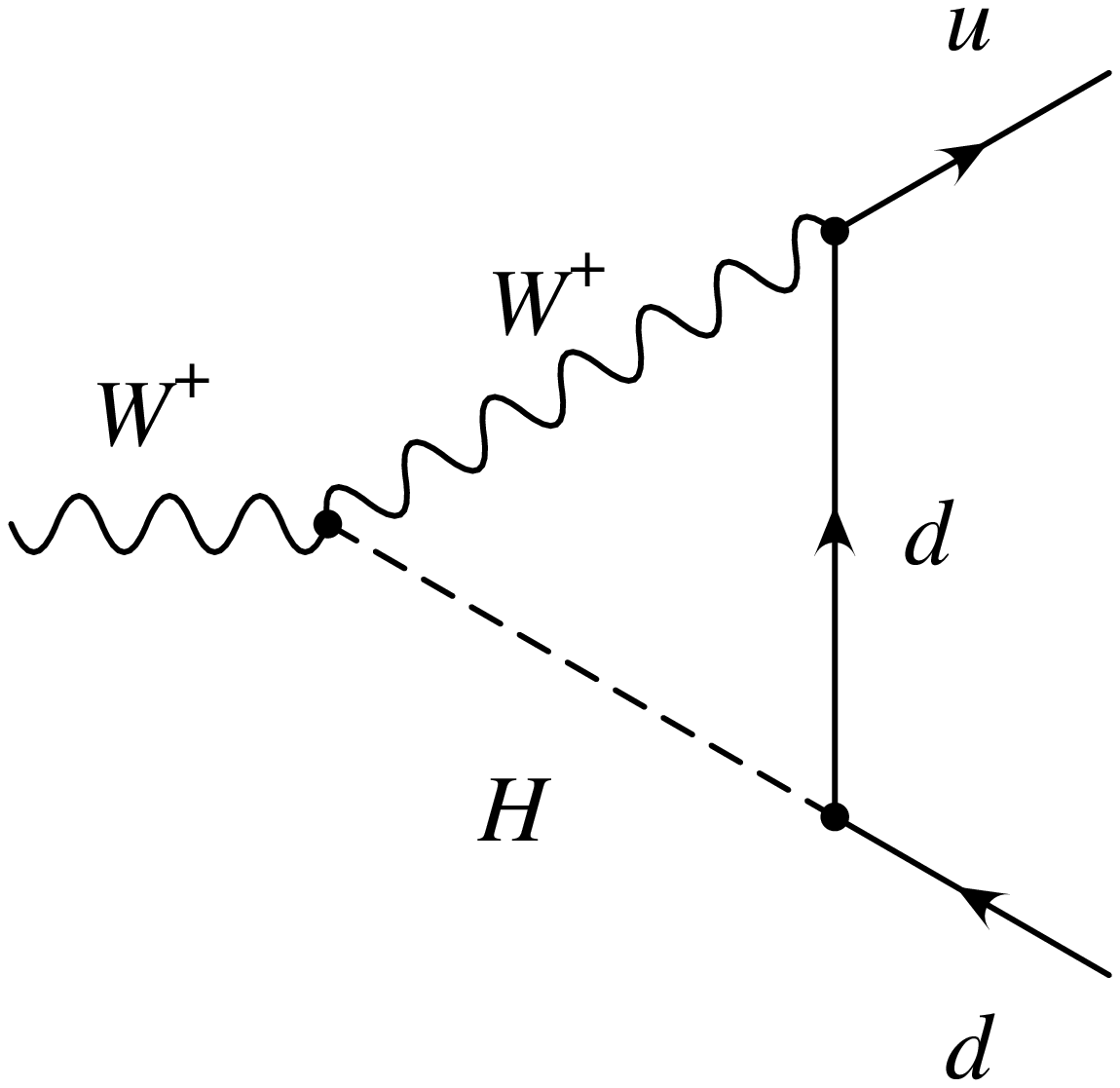,width=3.6cm}
    \epsfig{file=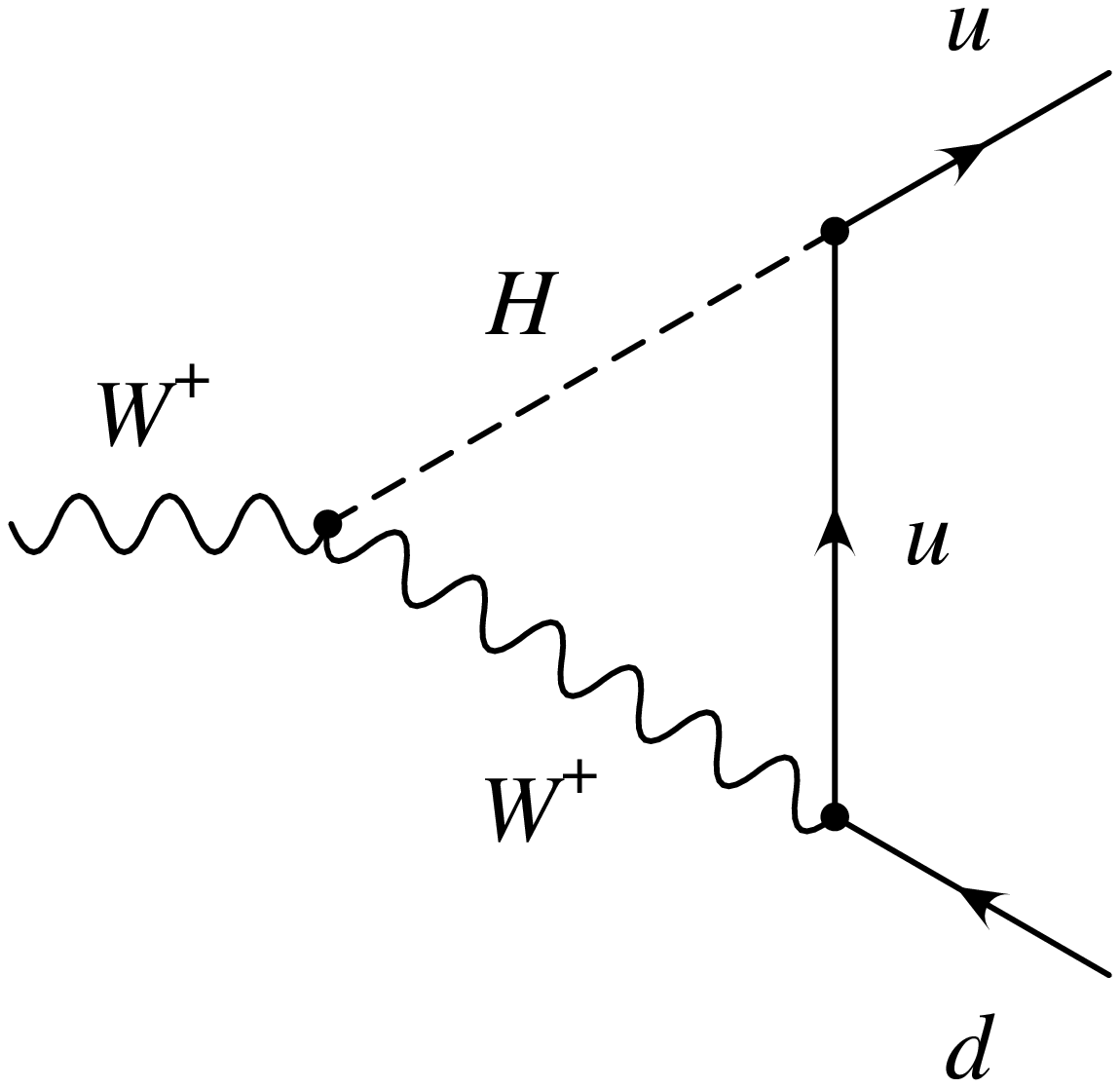,width=3.6cm}
    \epsfig{file=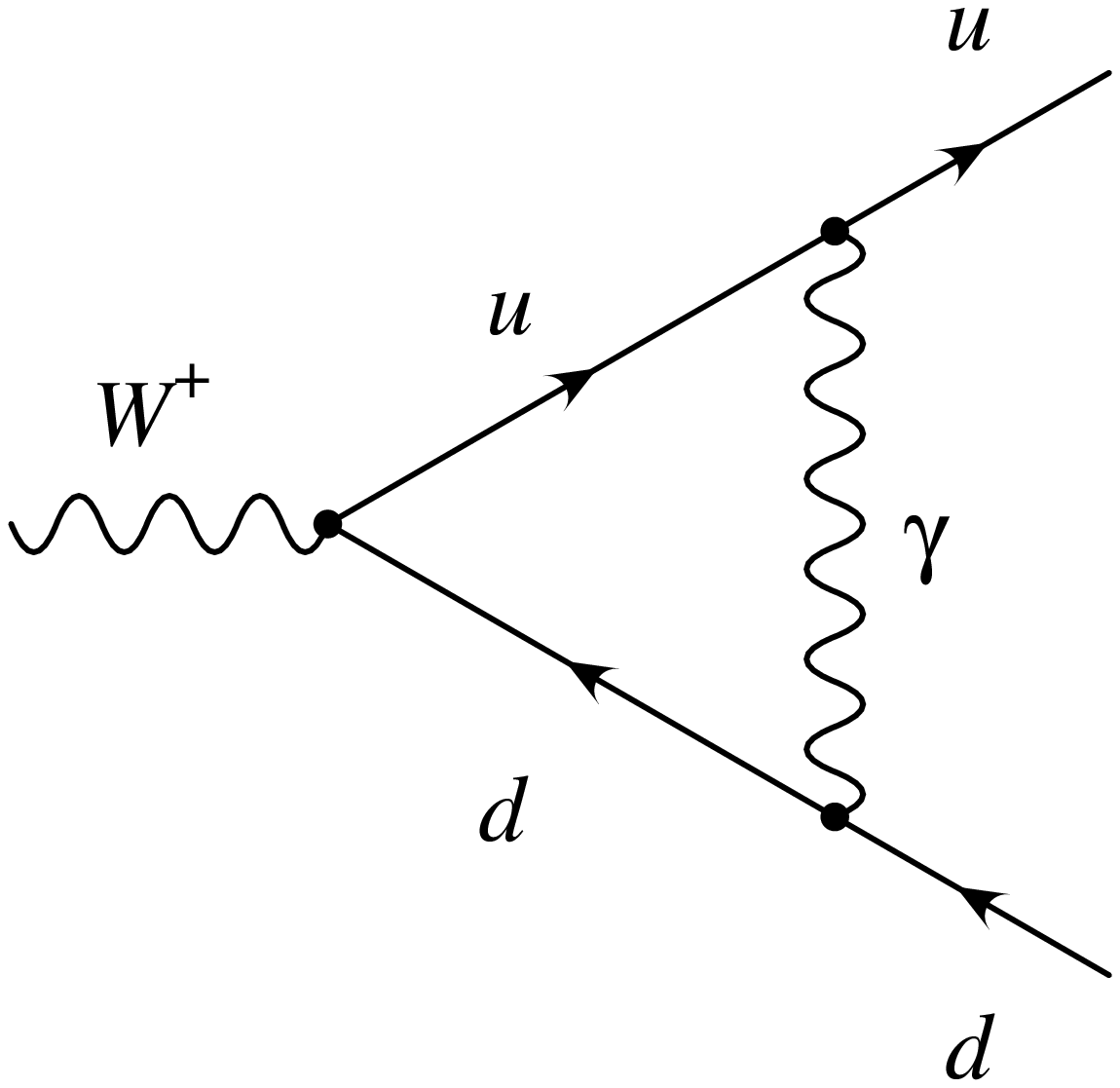,width=3.6cm}
    \epsfig{file=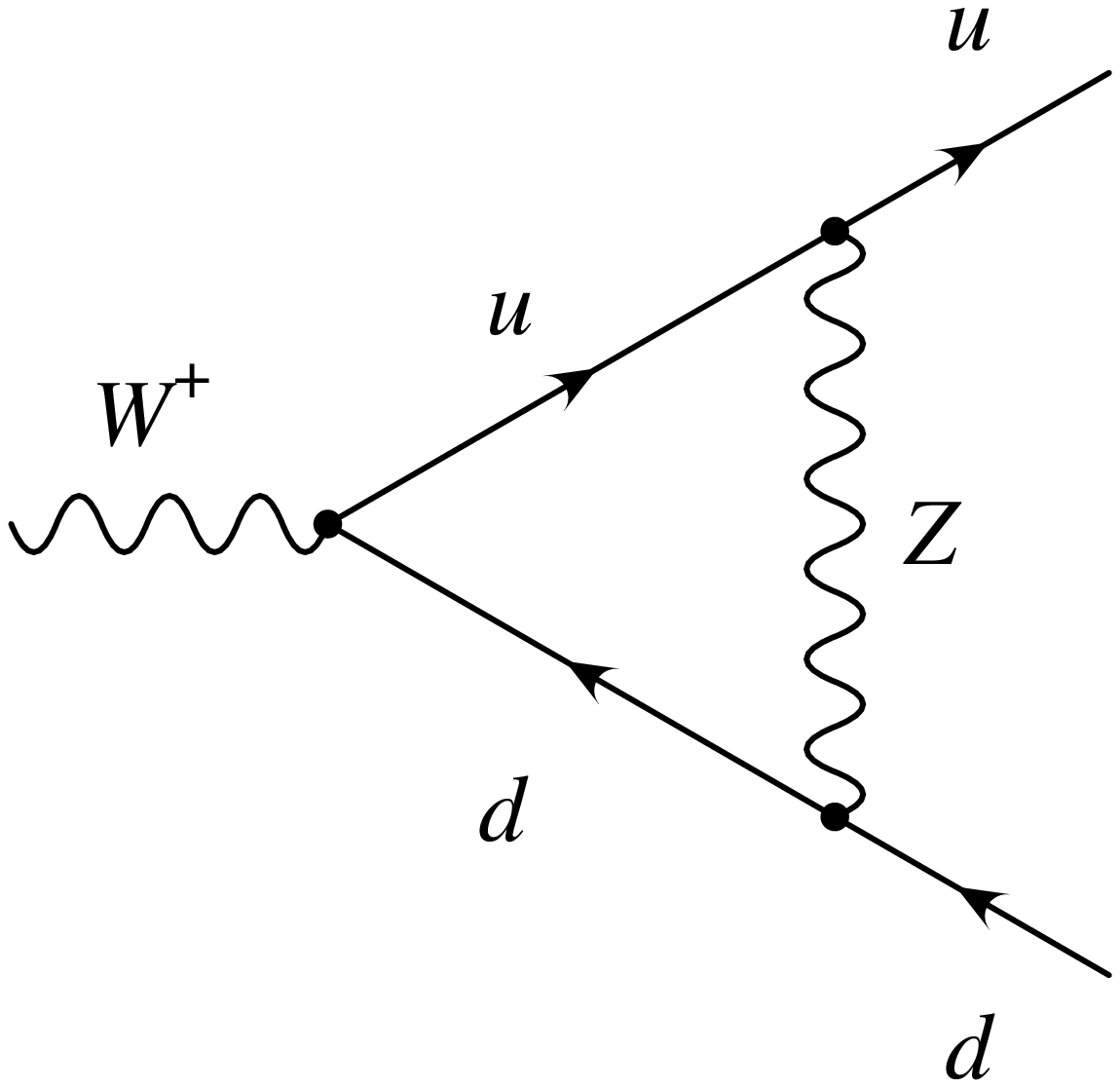,width=3.6cm}\\
    \epsfig{file=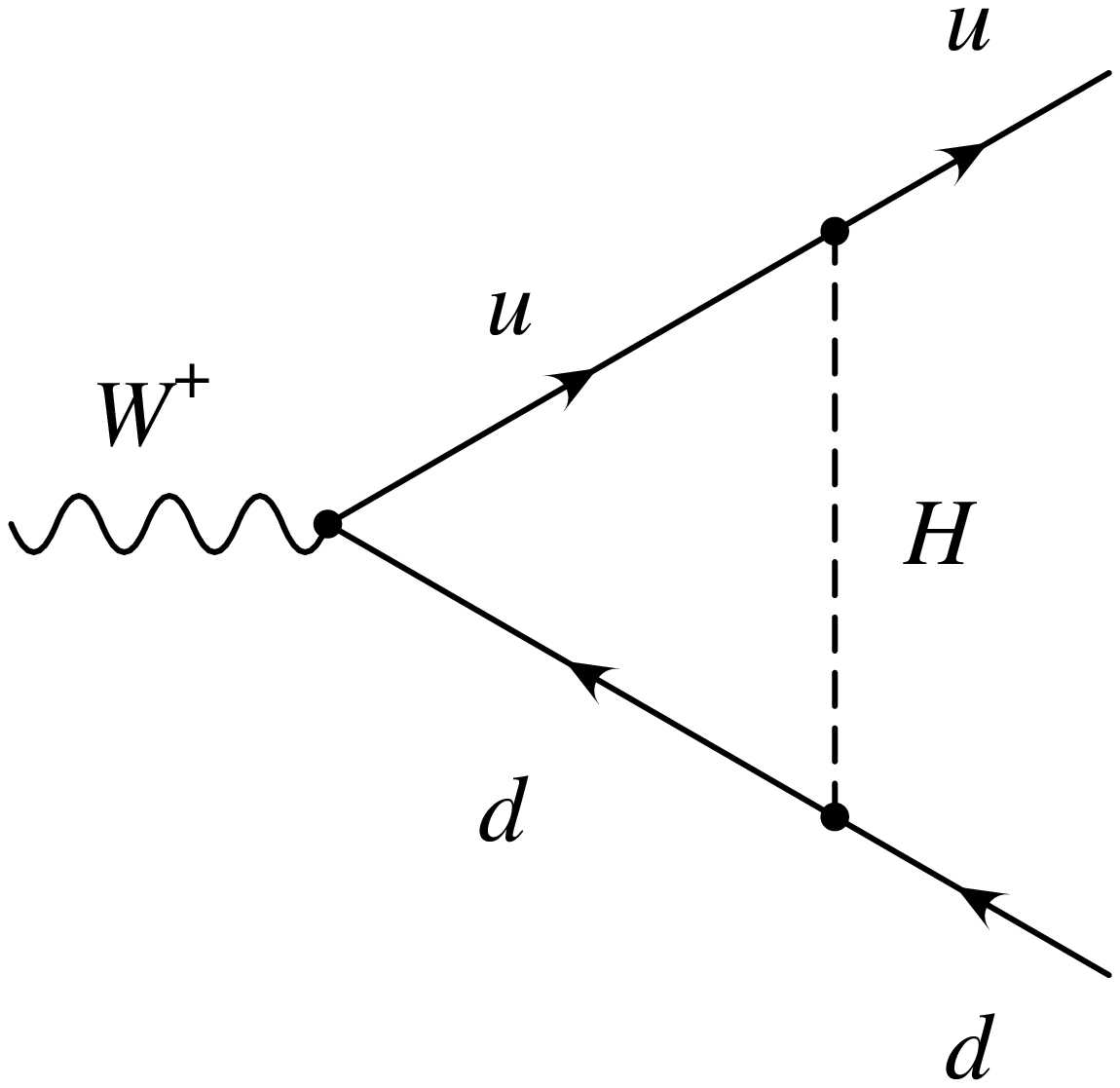,width=3.6cm}
    \epsfig{file=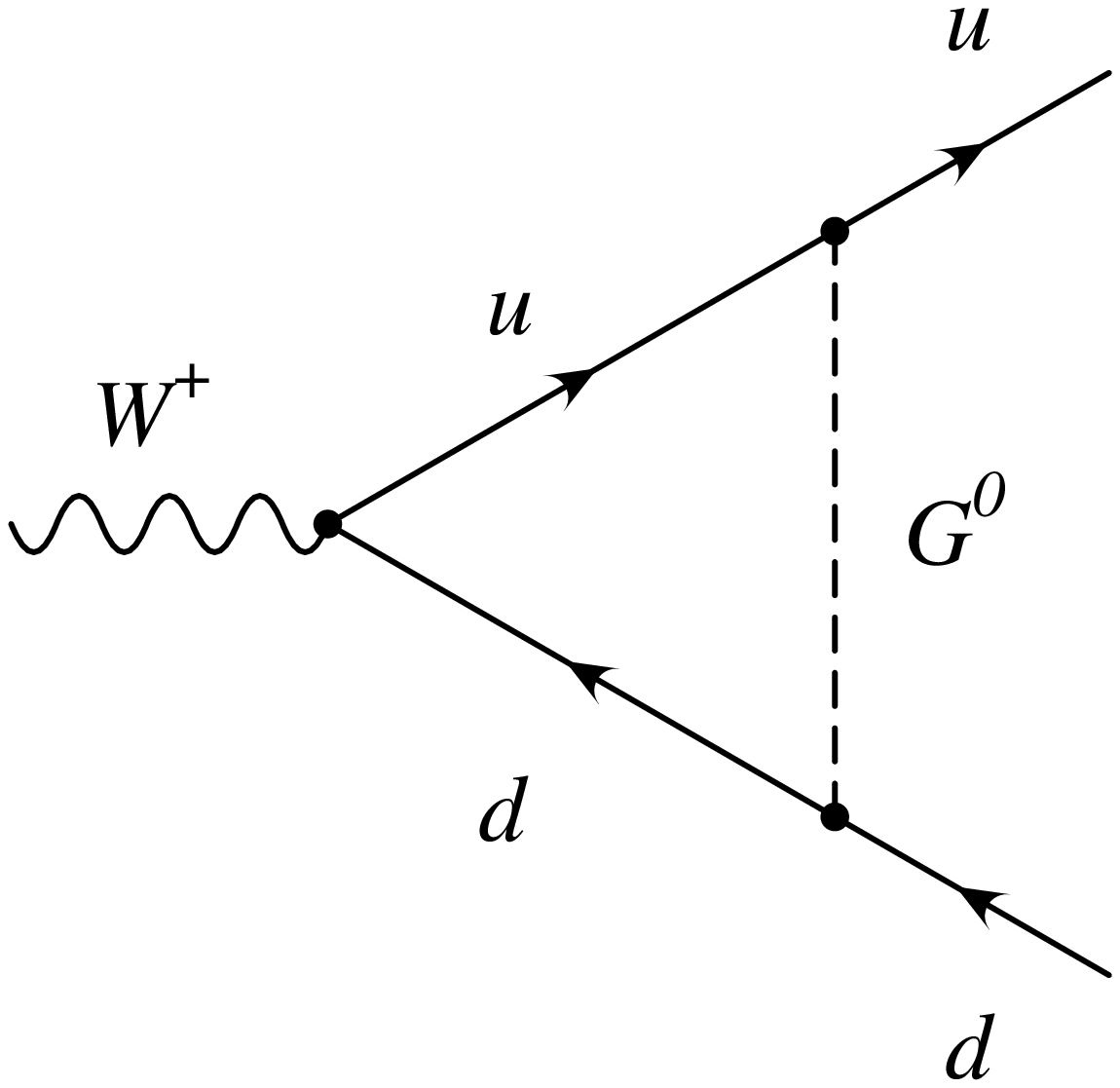,width=3.6cm}
    \epsfig{file=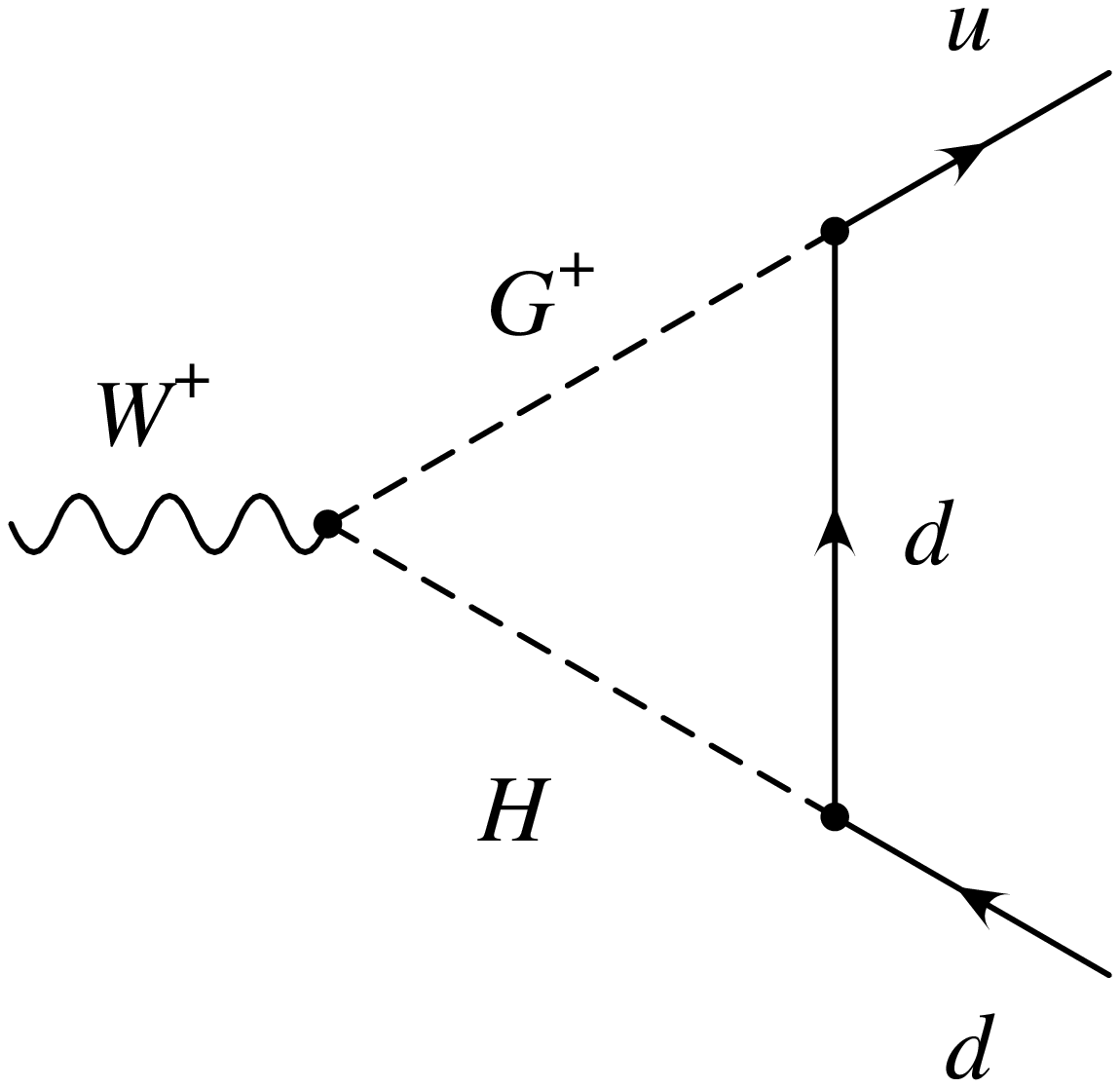,width=3.6cm}
    \epsfig{file=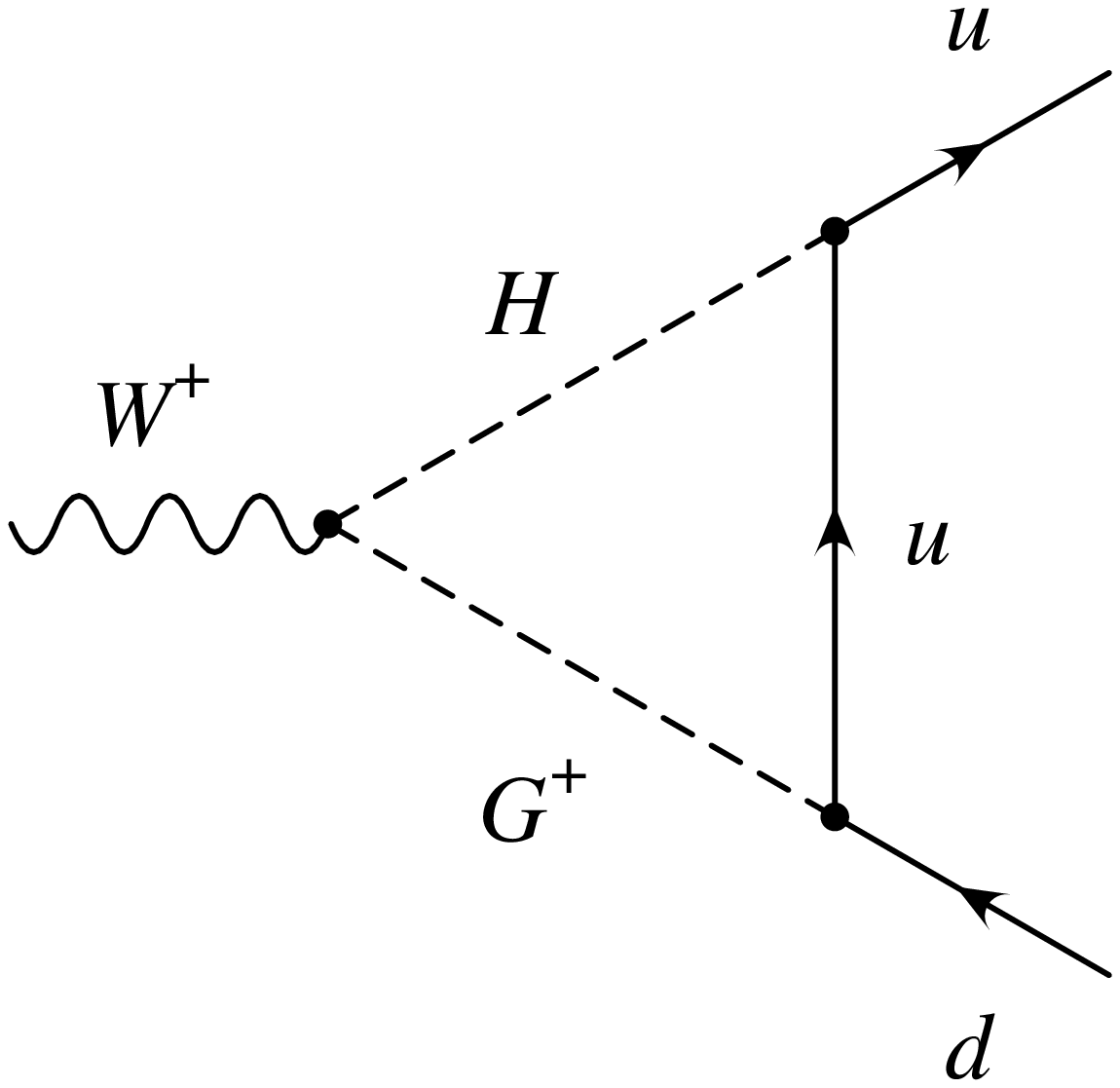,width=3.6cm}\\
    \epsfig{file=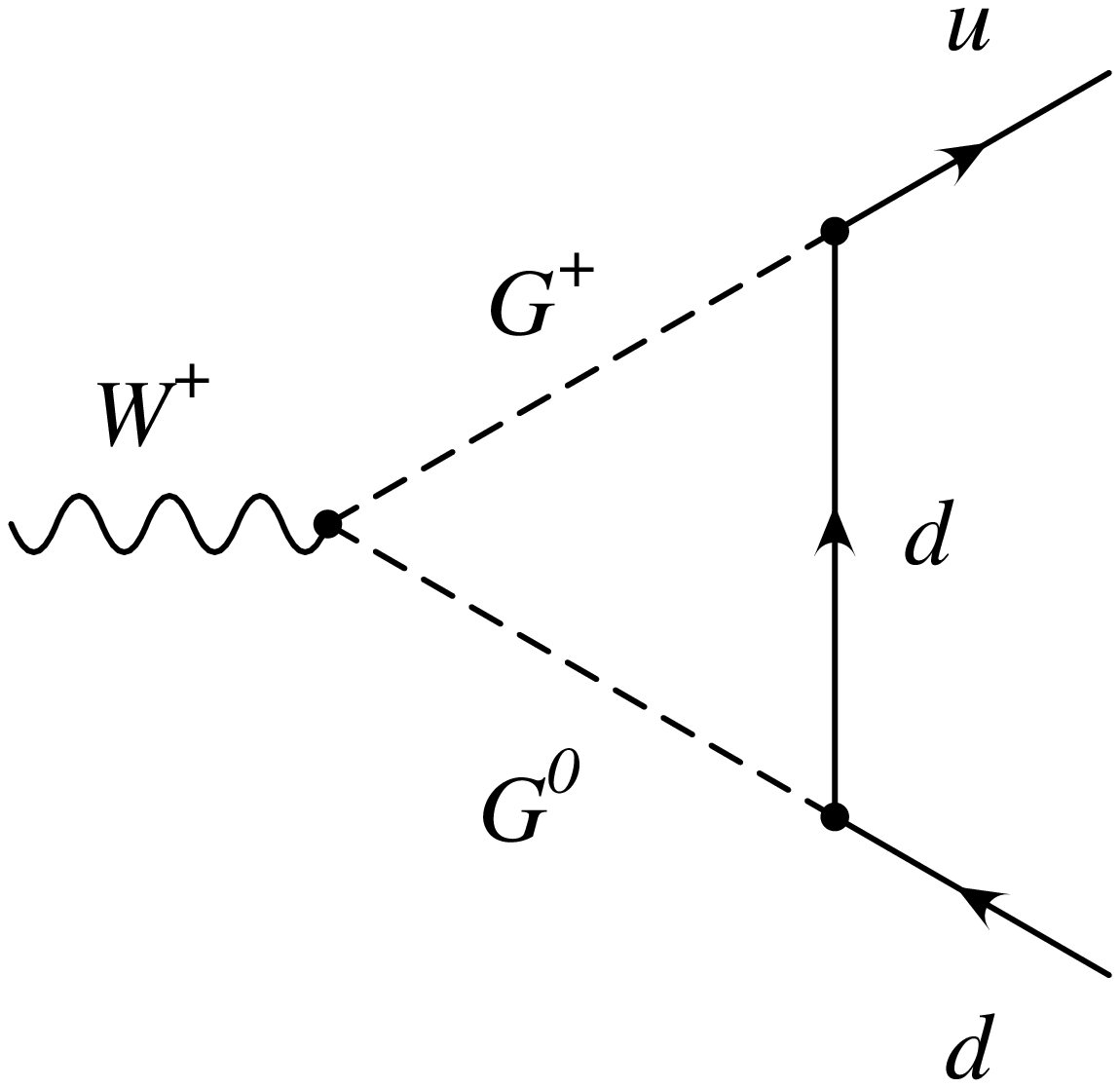,width=3.6cm}
    \epsfig{file=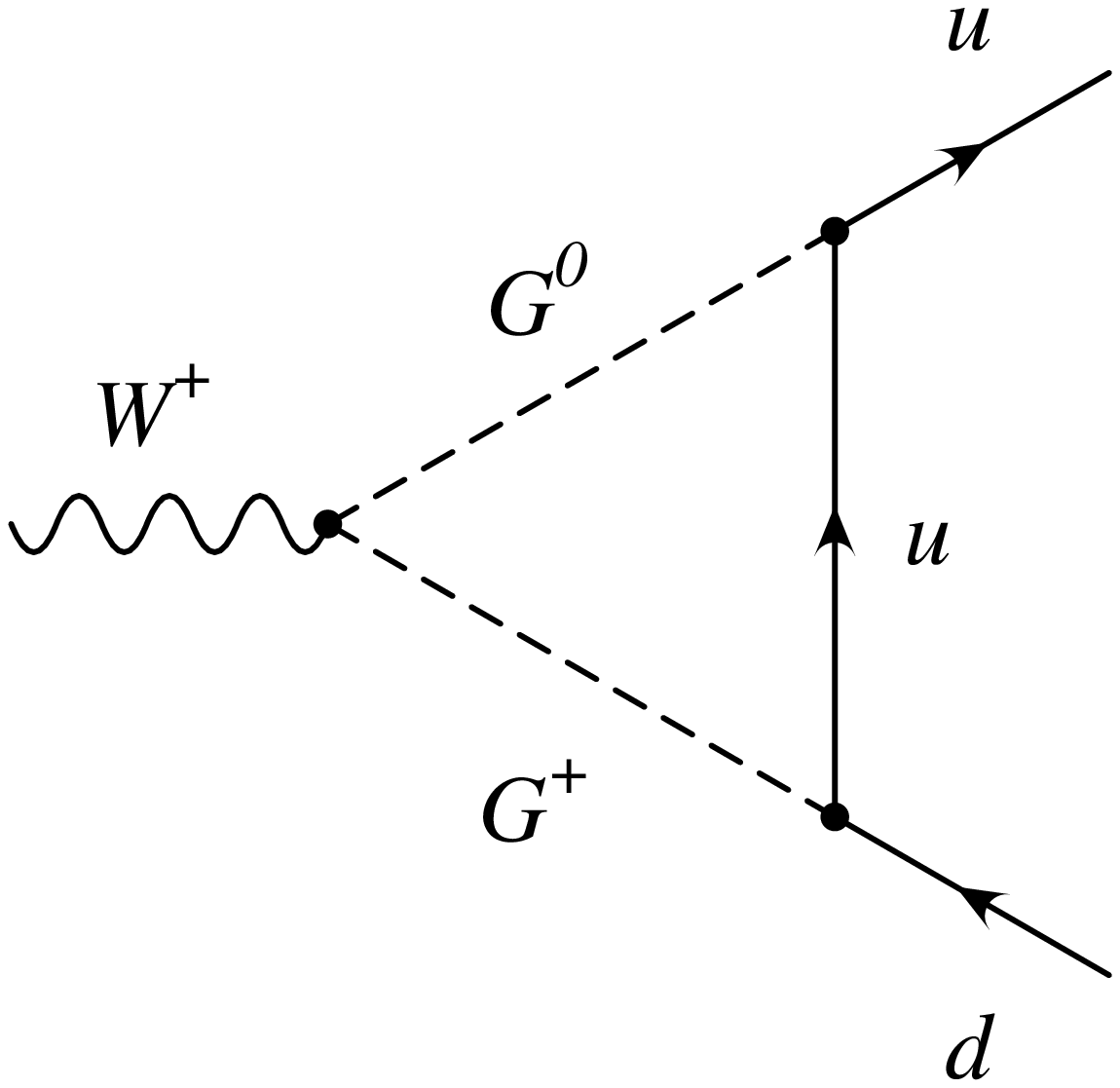,width=3.6cm}
    \epsfig{file=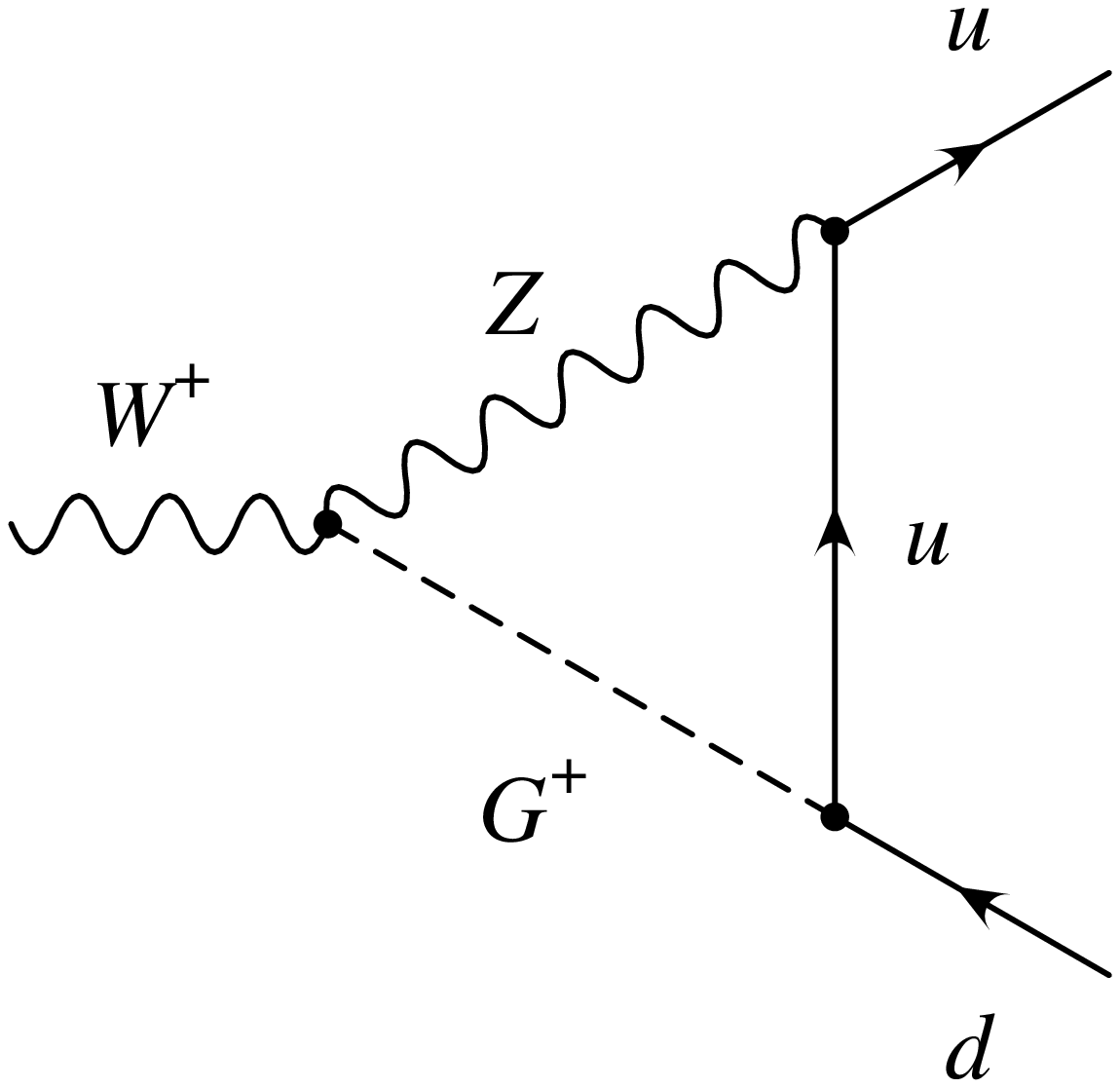,width=3.6cm}
    \epsfig{file=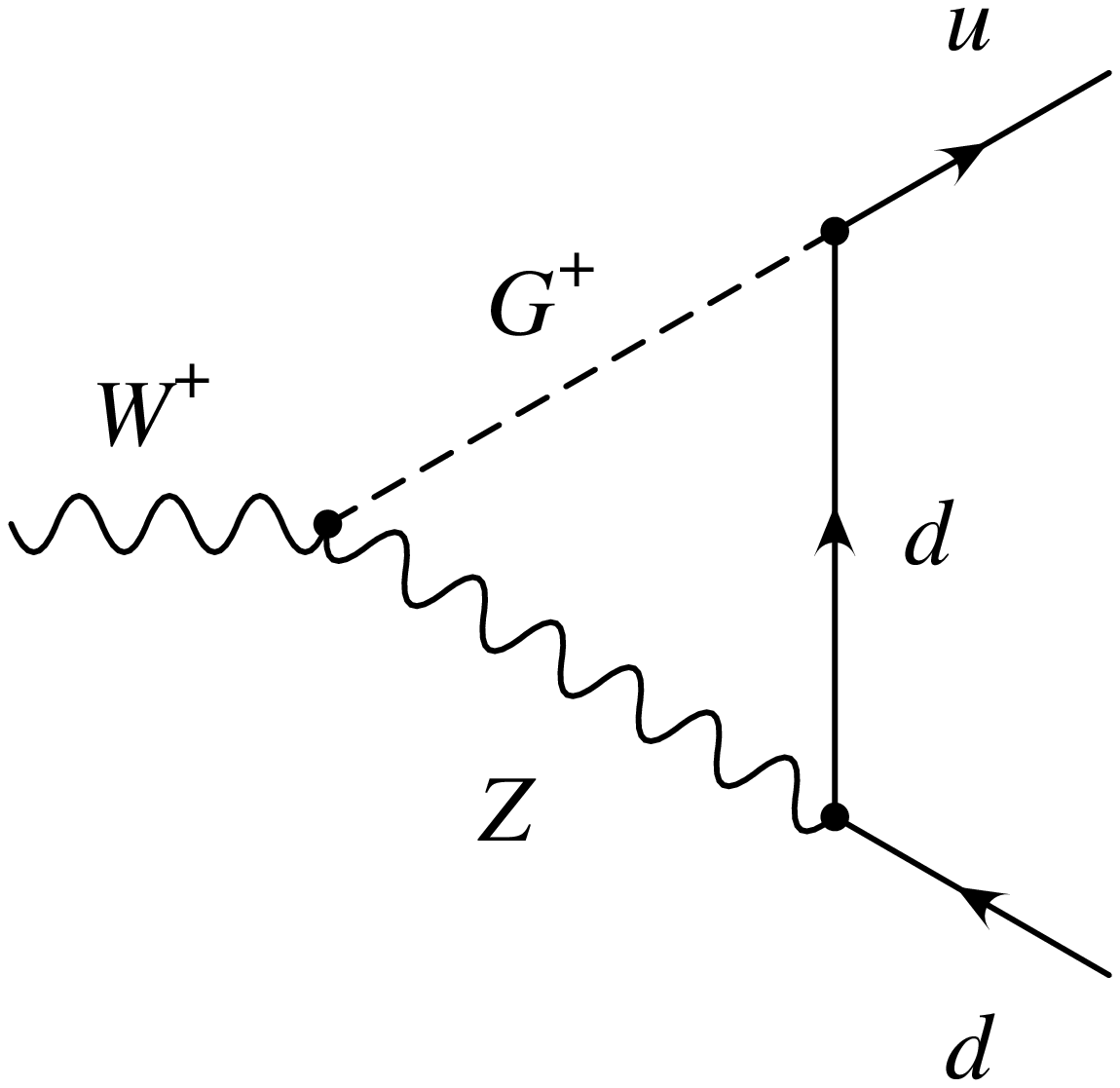,width=3.6cm}\\
    \epsfig{file=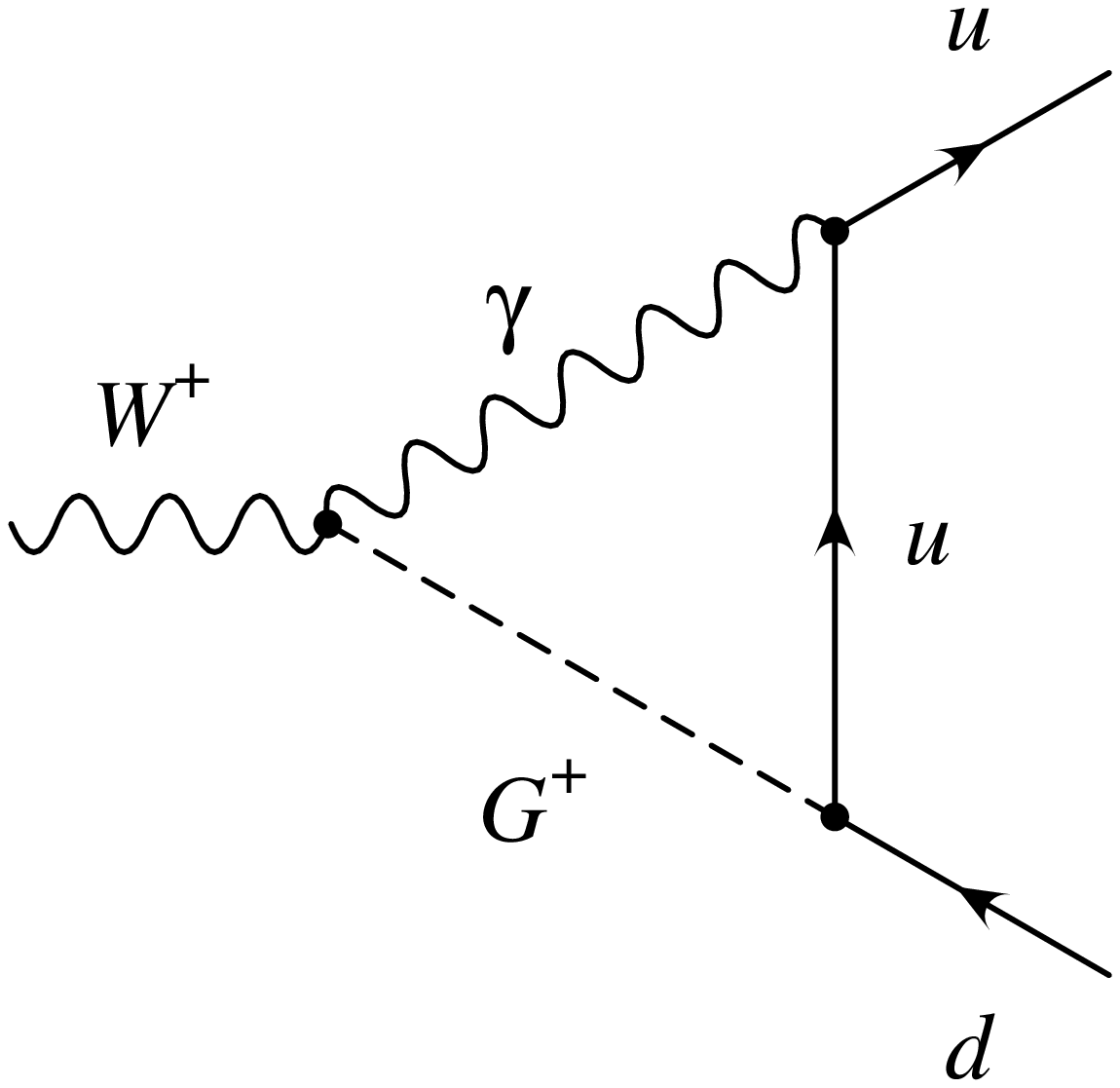,width=3.6cm}
    \epsfig{file=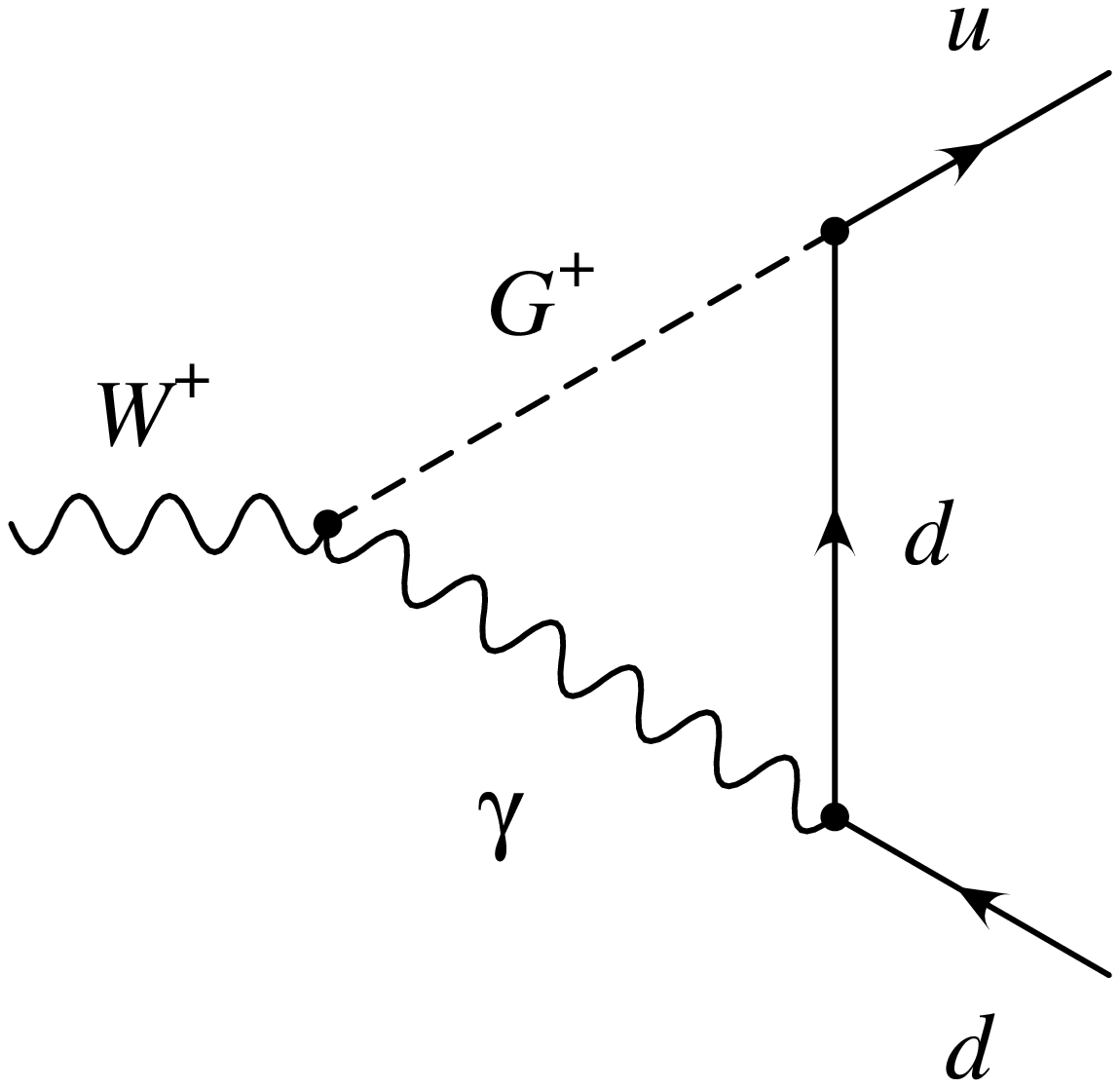,width=3.6cm}\\
   \caption{Irreducible electroweak one-loop diagrams for 
   $W^{+}\rightarrow u_i \bar{d}_j$}
  \end{center}
\end{figure}
\hspace{-3mm}Since each diagram in Fig.1 contains only one vertex of $W$ 
boson coupling to quarks, the form factors $F_{L,R}$ and $G_{L,R}$ are 
free of CKM matrix element. On the other hand, $F_R$ and $G_{L,R}$ are 
gauge independent \cite{c7,c9}.

The main idea of Ref.\cite{c9,c4} is to choose the CKM counterterm to 
make the amplitude $T_1$ in proportion to the fictitious amplitude of 
$W^{+}\rightarrow u_i \bar{d}_j$ in such a model that no quark-mixing
effect is introduced. The ratio is the CKM matrix element $V_{ij}$. As we know
if there is no quark-mixing effect, the amplitude $W^{+}\rightarrow u_i \bar{d}_j$ 
must be gauge independent and ultraviolet finite without introducing any CKM matrix renormalization. So such a renormalized result of $T_1$ is acceptable. 
Ref.\cite{c9} has suggested that such amplitude can be obtained from the 
amplitude $T_1$ by some modifications
\beq
  T_1\,=\,V_{ij}[A_L(F_L+\frac{\delta g}{g}+\half\delta Z_W +
  \half\delta \bar{Z}^{uL}_{ii[l]}+\half\delta Z^{dL}_{jj[l]})+
  A_R F_R+B_L G_L+B_R G_R] 
\eeq
where the subscript $[l]$ denotes the quantity is obtained by replacing CKM
matrix elements by unit matrix elements. Noted that the non-diagonal quark's 
WRC and the CKM counterterm has been removed. 

But in fact such a renormalized amplitude $T_1$ is ultraviolet divergent when
$i\not=j$. It is easy to calculate the ultraviolet divergent part of $T_1$ in 
Eq.(3), as shown below
$$ T_1 |_{UV-divergence}\,=\,\frac{\alpha V_{ij} \Delta}{32\pi M_W^2 s_W^2}
   (m_{d,i}^2-m_{d,j}^2+m_{u,j}^2-m_{u,i}^2) $$
with $\alpha$ the fine structure constant, $s_W$ the sine of the weak mixing
angle $\theta_W$, $m_{d,i}$ and $m_{d,j}$ the down-type quark masses, 
$m_{u,i}$ and $m_{u,j}$ the up-type quark masses, and 
$\Delta=2/(D-4)+\gamma_E-\ln(4\pi)+\ln(M^2_W/ \mu^2)$ (D is the space-time 
dimensionality, $\gamma_E$ is the Euler's constant and $\mu$ is an arbitrary energy 
scale). This result shows that when $i\not=j$ the decaying amplitude $T_1$ 
in Eq.(3) is ultraviolet divergent. 

We argue that such defect comes from the one-sided recognition of the difference
between the two case of having quark's mixing and not having quark's mixing. 
In the case of "not having quark's mixing", just like the case of leptons, no CKM 
matrix element is present at the fermion line which connects with the external fermion lines and only the fermions in a same generation can be present at one fermion line which connects with the external fermion lines in a Feynman diagram. Since in Eq.(3) we have recognized $u_i$ and $d_j$ as the "same generation" quarks, only the quarks $u_i$ and 
$d_j$ are permitted to be present at the inner fermion line which connects with the external fermion lines $u_i$ and $d_j$. Thus different from Eq.(3), the amplitude $T_1$ should be renormalized as follows:
\beq
  T_1\,=\,V_{ij}[A_L(F_L+\frac{\delta g}{g}+\half\delta Z_W +
  \half \delta \bar{Z}^{uL}_{ii[l]m_{d,i}\rightarrow m_{d,j}}+
  \half \delta Z^{dL}_{jj[l]m_{u,j}\rightarrow m_{u,i}})+A_R F_R+B_L G_L+B_R G_R] 
\eeq
This is because: 
\begin{enumerate}
\item the form factors $F_{L,R}$ and $G_{L,R}$ are free of CKM matrix element and only contain the contributions of the inner-line quarks $u_i$ and $d_j$, so don't need to 
be changed.
\item $\delta g$ is the counterterm of the coupling constant $g$. $\delta Z_W$ 
is the W boson's WRC. They all have nothing to do with fermion's property, so don't
need to be considered in this procedure.
\item we want to remove all of the other CKM matrix elements except for $V_{ij}$.
So the non-diagonal fermion's WRC in the terms 
$\sum_{k}\half\delta\bar{Z}^{uL}_{ik}V_{kj}+\sum_{k}\half V_{ik}\delta Z^{dL}_{kj}$ 
disappear and the diagonal fermion's WRC must be added an operator $[l]$ to remove the left CKM matrix elements.
\item Since only the quarks in the same generation of the external-line
quarks can contribute to the amplitude, the contribution of quark $d_i$ in 
$\delta\bar{Z}^{uL}_{ii[l]}$ and the contribution of quark $u_j$ in 
$\delta Z^{dL}_{jj[l]}$ (see Fig.2) must be changed to the contributions of quarks 
$d_j$ and $u_i$. This is realized by the operations $m_{d,i}\rightarrow m_{d,j}$ 
and $m_{u,j}\rightarrow m_{u,i}$. 
\end{enumerate}
Compared with Eq.(1) and (4) the CKM counterterm is
\beq
  \delta V_{ij}\,=\,-\half\sum_k[\delta \bar{Z}^{uL}_{ik}V_{kj}+
  V_{ik}\delta Z^{dL}_{kj}]+\half V_{ij}[
  \delta \bar{Z}^{uL}_{ii[l]m_{d,i}\rightarrow m_{d,j}}+
  \delta Z^{dL}_{jj[l]m_{u,j}\rightarrow m_{u,i}}]
\eeq
Our calculations have proved that this CKM counterterm is gauge independent and 
makes the physical amplitude $T_1$ in Eq.(4) ultraviolet convergent. 

Does this prescription keep the unitarity of the CKM matrix? Here we will 
do some analytical calculations to show in what degree $\delta V$ satisfies 
the unitary condition. We have split the bare CKM matrix element $V^0_{ij}$ 
into $V^0_{ij}=V_{ij}+\delta V_{ij}$ and keep the unitarity of the renormalized 
CKM matrix $V$ in the previous calculations. So the unitarity condition of the 
bare CKM matrix $V^0$ requires
\beq
  \sum_k(\delta V_{ik}V^{\ast}_{jk}+V_{ik}\delta V^{\ast}_{jk})\,=\,
  \sum_k(V^{\ast}_{ki}\delta V_{kj}+\delta V^{\ast}_{ki}V_{kj})\,=\,0
\eeq
At one-loop level only four diagrams have contributions to the CKM counterterm 
in Eq.(5), as shown in Fig.2.
\begin{figure}[tbh]
\begin{center}
  \epsfig{file=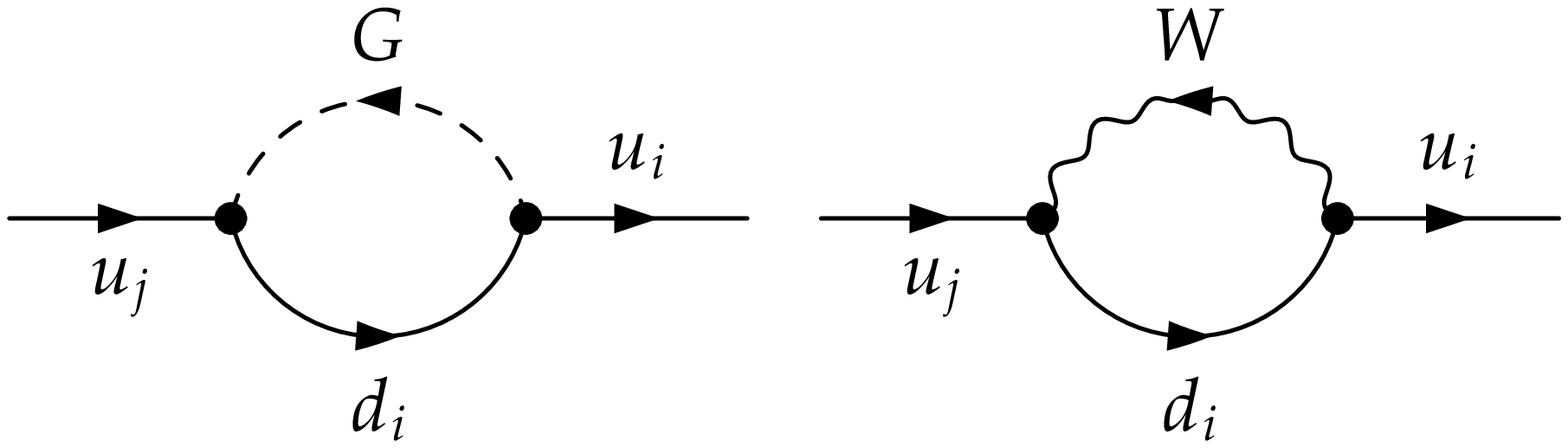, width=7.2cm} 
  \epsfig{file=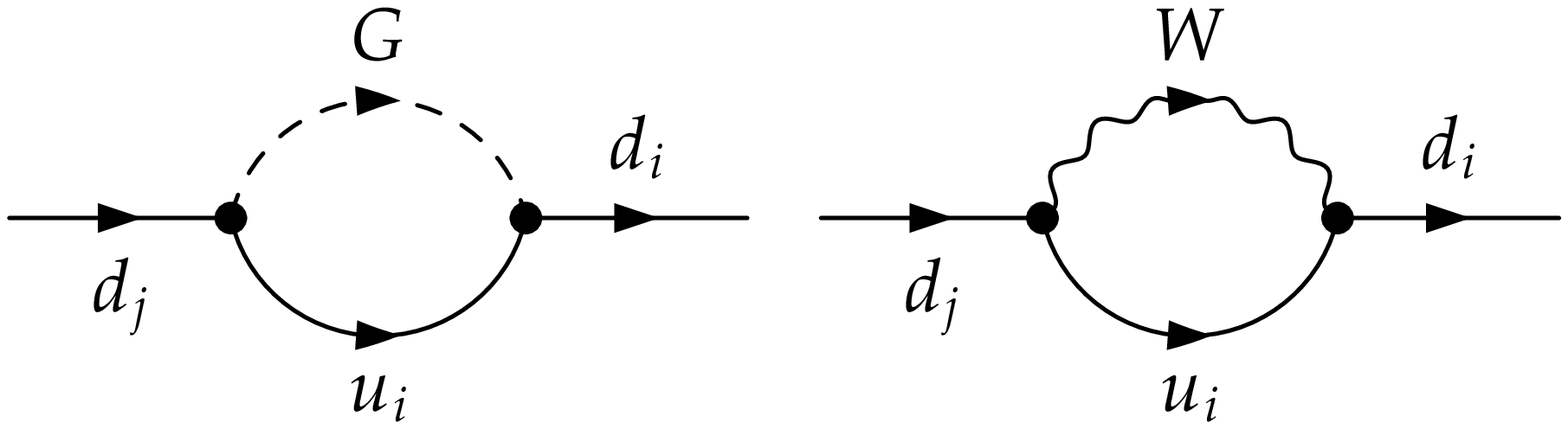, width=7.2cm}
  \caption{Quark's self-energy diagrams that contribute to the CKM counterterm
  in Eq.(5).}
\end{center} \end{figure}
\hspace{-3mm}We have used the mathematica package {\em FeynArts} \cite{c11} to 
draw the Feynman diagrams and generate the corresponding Feynman amplitudes, 
and used the mathematica package {\em FeynCalc} \cite{c12} to calculate these 
Feynman amplitudes. It is easy to get the analytical result of $\delta V_{ij}$ 
since the quark's self-energy functions at one-loop level are simple. In order 
to check the unitary condition of Eq.(6) analytically, we use the Taylor's 
series: $(m^2_{quark}/M^2_W)^n$, to expand $\delta V_{ij}$. The one- and 
two-order results are shown below:
\beqa
  \delta V^{(1)}_{ij}\,=\,&&\frac{\alpha(6\Delta-11)}{128\pi M^2_W s^2_W}
  [-\frac{2\sum_{k,l\not=j}m_{d,j}m^2_{u,k}V_{il}V^{\ast}_{kl}V_{kj}}{m_{d,l}-
  m_{d,j}}+\frac{2\sum_{k,l}m_{d,j}m^2_{u,k}V_{il}V^{\ast}_{kl}V_{kj}}{m_{d,l}+
  m_{d,j}}-\frac{2\sum_{k\not=i,l}m_{u,i}m^2_{d,l}V_{il}V^{\ast}_{kl}V_{kj}}
  {m_{u,k}-m_{u,i}} \nonumber \\ &&+  
  \frac{2\sum_{k,l}m_{u,i}m^2_{d,l}V_{il}V^{\ast}_{kl}V_{kj}}{m_{u,k}+
  m_{u,i}}+V_{ij}(\sum_k V_{ik}V^{\ast}_{ik}m^2_{d,k}+
  \sum_k V_{kj}V^{\ast}_{kj}m^2_{u,k}-2 m^2_{d,j}-2 m^2_{u,i})]\,,
\eeqa
\beqa
  \delta V^{(2)}_{ij}\,=\,&&\frac{\alpha}{64\pi M^4_W s^2_W}[V_{ij}
  (12 \ln\frac{m^2_{d,j}}{M^2_W} m^4_{d,j}+6 m^4_{d,j}+
  12 \ln\frac{m^2_{u,i}}{M^2_W} m^4_{u,i}+6 m^4_{u,i}+32 m^2_{d,j}m^2_{u,i} 
  \nonumber \\ &&-
  \sum_k V_{ik}V^{\ast}_{ik}(6 \ln\frac{m^2_{d,k}}{M^2_W} m^4_{d,k}+3 m^4_{d,k}+
  12 m^2_{u,i}m^2_{d,k})-\sum_k V_{kj}V^{\ast}_{kj}
  (6 \ln\frac{m^2_{u,k}}{M^2_W}m^4_{u,k}+3 m^4_{u,k}+12 m^2_{d,j}m^2_{u,k})) 
  \nonumber \\ &&+
  \frac{2}{m_{d,l}-m_{d,j}}\sum_{k,l\not=j}m_{d,j}m^2_{u,k}(4 m^2_{d,j}+
  6 \ln\frac{m^2_{u,k}}{M^2_W} m^2_{u,k}+3 m^2_{u,k})V_{il}V^{\ast}_{kl}V_{kj} 
  \nonumber \\ &&-
  \frac{2}{m_{d,l}+m_{d,j}}\sum_{k,l}m_{d,j}m^2_{u,k}
  (4 m^2_{d,j}+6 \ln\frac{m^2_{u,k}}{M^2_W} m^2_{u,k}+3 m^2_{u,k})
  V_{il}V^{\ast}_{kl}V_{kj} \nonumber \\ &&+
  \frac{2}{m_{u,k}-m_{u,i}}\sum_{k\not=i,l}m_{u,i}m^2_{d,l}(4 m^2_{u,i}+
  6 \ln\frac{m^2_{d,l}}{M^2_W} m^2_{d,l}+3 m^2_{d,l})V_{il}V^{\ast}_{kl}V_{kj} 
  \nonumber \\ &&-
  \frac{2}{m_{u,k}+m_{u,i}}\sum_{k,l}m_{u,i}m^2_{d,l}(4 m^2_{u,i}+
  6 \ln\frac{m^2_{d,l}}{M^2_W} m^2_{d,l}+3 m^2_{d,l})V_{il}V^{\ast}_{kl}V_{kj}]\,.
\eeqa
where the superscript $(1)$ and $(2)$ denote the one and two order results
of $\delta V_{ij}$ about the series $m^2_{quark}/M^2_W$. The $R_{\xi}$-gauge 
has been used. Noted that $\delta V^{(1)}_{ij}$ and $\delta V^{(2)}_{ij}$ are
both gauge independent. Replacing $\delta V$ with $\delta V^{(1)}+\delta V^{(2)}$
in Eq.(6), we find it satisfies the unitary condition.  

But when we consider the three-order result of $\delta V$ about the series
$m^2_{quark}/M^2_W$, we find that
\beqa
  \sum_k(\delta V^{(3)\ast}_{ki}V_{kj}+V^{\ast}_{ki}\delta V^{(3)}_{kj})\,=\,&&
  \frac{9\alpha}{128\pi M^6_W s^2_W}[\sum_k m^2_{u,k}(m^4_{d,i}-
  2 m^2_{d,j}m^2_{d,i}+m^4_{d,j}+m^2_{d,i}m^2_{u,k}+
  m^2_{d,j}m^2_{u,k})V^{\ast}_{ki}V_{kj} \nonumber \\ &&-
  2\sum_{k,l,n}m^2_{u,k}m^2_{d,l}m^2_{u,n}V^{\ast}_{ki}V_{kl}V^{\ast}_{nl}V_{nj}]
  \,\not=\,0
\eeqa
This result shows that $\delta V$ doesn't comply with the unitary condition. But
from this result we can see that the deviation of $\delta V$ from the unitary 
condition is very small, since the quark's masses are very small compared with 
$M_W$ (except for $m_t$). Calculating till to five-order result of $\delta V$ about 
the series $m^2_{quark}/M^2_W$, we find the largest deviation of 
$\delta V^{\dagger}V+V^{\dagger}\delta V$ from $0$ is proportional to 
$\alpha|V_{3i\not=3}|m^2_b m^8_t/(s_W^2 M^{10}_W) \sim 10^{-7}$, which is very 
small compared with the present measurement precision of the CKM matrix elements.
Thus in actual calculations we can use Eq.(5) as the definition of the CKM 
counterterm. Comparing with the prescription in Ref.\cite{c4} one can see that 
our prescription is very simple and suitable for actual calculations.

\section{Relationship between the Unitarity and the Gauge Independence of the 
CKM Matrix}

It has been proved that any physical parameter's counterterm must be gauge
independent \cite{c13} if they don't break up the theory structure. So  
the CKM matrix counterterm is also gauge independent \cite{c7} if it doesn't
break up the standard model's structure, i.e. the unitarity of the CKM matrix. 
There has been proposed a negative proposition that in order to keep the gauge independence of the CKM counterterm the CKM renormalization prescription must keep 
the unitarity of the CKM matrix \cite{c7,c4}. Here we want to give a Proof for this hypothesis by analytical calculations.

In general one will encounter a question: if the CKM counterterms in
lower-loop levels are gauge independent but don't satisfy the unitary 
condition, will they change the gauge independence of the CKM counterterms
in the higher-loop levels? If the answer is "yes", the above hypothesis must 
be true. In order to study this question we consider the most simple
case: the effect of the unitarity of the one-loop CKM counterterm to 
the gauge independence of the two-loop CKM counterterm. 

In order to elaborate this problem clearly we express the amplitude of
$W^{+}\rightarrow u_i \bar{d}_j$ as
\beq
  T(V^0)\,=\,T(V+\delta V)\,=\,T(V)+T^{\prime}(V)\delta V+
  \half T^{\prime\prime}(V)(\delta V)^2+\cdot\cdot\cdot
\eeq
where the superscript $\prime$ denotes the partial derivative with respect to
the CKM matrix elements. Since the CKM counterterm has been written out 
apparently, the amplitude $T$ on the right-hand side of Eq.(10) doesn't 
contain CKM counterterm any more. To two-loop level, this equation becomes
\beq
  T_2(V^0)\,=\,T_2(V)+T^{\prime}_1(V)\delta V_1+\delta V_2 A_L
\eeq
where the subscripts $1$ and $2$ denote the 1-loop and 2-loop results. 
Since $T_2(V^0)$ must be gauge independent, the gauge dependence of 
$\delta V_2$ is determined by the gauge dependence of $T^{\prime}_1(V)\delta V_1$
and $T_2(V)$. In the following we will firstly prove that $T_2(V)$ is gauge 
independent. 

Using the facts that the one-loop formfactors $F_R$ and $G_{L,R}$ are gauge
independent and don't contain CKM matrix element, and the terms in the first
bracket of Eq.(1) are gauge independent \cite{c7}, one get
\beqa
  T^{\prime}_1(V)\delta V_1 |_{\xi}=&&[-\frac{\delta V_{ij}}{2 V_{ij}}
  (\sum_{k\not=i}\delta \bar{Z}^{uL}_{ik}V_{kj}+
  \sum_{k\not=j}V_{ik}\delta Z^{dL}_{kj})+\frac{V_{ij}}{2}\sum_{k,l}
  (\frac{2}{g}\frac{d(\delta g)}{d V_{kl}}+\frac{d(\delta Z_W)}{d V_{kl}}+
  \frac{d(\delta \bar{Z}^{uL}_{ii})}{d V_{kl}}+
  \frac{d(\delta Z^{dL}_{jj})}{d V_{kl}})\delta V_{kl} \nonumber \\ &&+ 
  \half(\sum_{k\not=i}\delta \bar{Z}^{uL}_{ik}\delta V_{kj}+
  \sum_{l,m,k\not=i}\frac{d(\delta \bar{Z}^{uL}_{ik})}{d V_{lm}}\delta V_{lm}V_{kj}
  +\sum_{k\not=j}\delta V_{ik}\delta Z^{dL}_{kj}+
  \sum_{l,m,k\not=j}V_{ik}\frac{d(\delta Z^{dL}_{kj})}{d V_{lm}}\delta V_{lm})]A_L
\eeqa
where the subscript $1$ of $\delta V_1$ on the right-hand side of Eq.(12)
has been omitted and the subscript $\xi$ on the left-hand side of Eq.(12)
denotes the gauge-dependent part of the quantity. Omitting the imaginary 
part of the quark's self-energy functions, one obtain
\beqa
  T^{\prime}_1(V)\delta V_1 &&|_{\xi}=\frac{\alpha A_L}{32\pi M^2_W s^2_W m^2_{d,j}}
  \sum_{k,l}(\delta V_{il}V^{\ast}_{kl}+V_{il}\delta V^{\ast}_{kl})V_{kj}[-
  \xi^2_W M_W^4\ln\frac{m_{u,k}^2}{M_W^2}+
  2\xi_W m_{d,j}^2 M_W^2\ln\frac{m_{u,k}^2}{M_W^2}-m_{u,k}^4\ln\xi_W 
  \nonumber \\ &&+
  2 m_{d,j}^2 m_{u,k}^2\ln\xi_W-2\sqrt{-\xi_W^2 M_W^4+2\xi_W m_{d,j}^2 M_W^2
  +2\xi_W m_{u,k}^2 M_W^2-m_{d,j}^4-m_{u,k}^4+2 m_{d,j}^2 m_{u,k}^2} 
  \nonumber \\ &&\times
  (\xi_W M_W^2-m_{d,j}^2+m_{u,k}^2)\arctan\frac{\sqrt{-\xi_W^2 M_W^4+
  2\xi_W m_{d,j}^2 M_W^2+2\xi_W m_{u,k}^2 M_W^2-m_{d,j}^4-m_{u,k}^4+
  2 m_{d,j}^2 m_{u,k}^2}}{-\xi_W M_W^2+m_{d,j}^2-m_{u,k}^2}] 
  \nonumber \\ &&+
  \frac{\alpha A_L}{32\pi M^2_W s^2_W m^2_{u,i}}\sum_{k,l}
  (V^{\ast}_{lk}\delta V_{lj}+\delta V^{\ast}_{lk}V_{lj})V_{ik}[-
  \xi^2_W M_W^4\ln\frac{m_{d,k}^2}{M_W^2}+
  2\xi_W m_{u,i}^2 M_W^2\ln\frac{m_{d,k}^2}{M_W^2}-m_{d,k}^4\ln\xi_W
  \nonumber \\ &&+
  2 m_{u,i}^2 m_{d,k}^2\ln\xi_W -2\sqrt{-\xi_W^2 M_W^4+2\xi_W m_{u,i}^2 M_W^2
  +2\xi_W m_{d,k}^2 M_W^2-m_{u,i}^4-m_{d,k}^4+2 m_{u,i}^2 m_{d,k}^2} 
  \nonumber \\ &&\times
  (\xi_W M_W^2-m_{u,i}^2+m_{d,k}^2)\arctan\frac{\sqrt{-\xi_W^2 M_W^4+
  2\xi_W m_{u,i}^2 M_W^2+2\xi_W m_{d,k}^2 M_W^2-m_{u,i}^4-m_{d,k}^4+
  2 m_{u,i}^2 m_{d,k}^2}}{-\xi_W M_W^2+m_{u,i}^2-m_{d,k}^2}] \nonumber \\
\eeqa
with $\xi_W$ the $W$ boson gauge parameter. It is easy to see that when
$\delta V_1$ satisfies the unitary condition of Eq.(6), 
$T^{\prime}_1(V)\delta V_1$ is gauge independent. By the precondition that the
CKM counterterm must be gauge independent if it doesn't break up the unitarity 
of the CKM matrix, we conclude that $T_2(V)$ is gauge independent. Because in 
this case both $\delta V_2$ and $T^{\prime}_1(V)\delta V_1$ are gauge independent, 
the gauge independence of the right-hand side of Eq.(11) yields $T_2(V)$ is also 
gauge independent.

Since $T_2(V)$ is free of CKM matrix counterterms (see above), it must be
gauge independent in spite of whether $\delta V_1$ satisfies the unitarity 
condition. According to Eq.(13), if $\delta V_1$ doesn't satisfy the unitary 
condition, $T^{\prime}_1(V)\delta V_1$ will be gauge dependent, 
thus $\delta V_2$ must be gauge dependent in order to make the right-hand side of 
Eq.(11) gauge independent. So we have proved the conclusion that if the one-loop CKM 
counterterm doesn't keep the unitarity of the CKM matrix the two-loop 
CKM counterterm must be gauge dependent. This conclusion is a strong Proof 
for the hypothesis that in order to keep the gauge independence 
of the CKM counterterm the CKM renormalization prescription must keep 
the unitarity of the CKM matrix.

\section{Conclusion}

In summary, We have investigated the present CKM matrix renormalization 
prescriptions and found the prescriptions proposed in Ref.\cite{c9,c4}
have some defect - either making the physical amplitude involving quark's 
mixing ultraviolet divergent in non-diagonal case or too complex. So we propose a new 
prescription which can make the physical amplitude involving quark's 
mixing ultraviolet finite and gauge independent and keeps the unitarity 
of the CKM matrix at a very high precision. The most important property
is our prescription is very simple compared with the previous prescriptions, 
so is suitable for actual calculations. On the other hand we have 
studied the relationship between the unitarity and the gauge independence 
of the CKM matrix. Through analytical calculations we have given a strong Proof for 
the important hypothesis that in order to keep the gauge independence of 
the CKM matrix the CKM renormalization prescription must keep the 
unitarity of the CKM matrix.

\vspace{5mm} {\bf \Large Acknowledgments} \vspace{2mm} 

The author thanks professor Xiao-Yuan Li for his useful guidance and Dr. Hu 
qingyuan for his sincerely help (in my life). The author also thanks Prof.
B. A. Kniehl very much for his valuable criticism.

\end{document}